# First principle study of crystal growth morphology: An application to crystalline urea

M. K. Singh[*]

*Laser Materials Development & Devices Division,
Raja Ramanna Centre for Advanced Technology,
Indore (M.P)*

The growth morphology and the effects of surface reconstruction/relaxation of the habit faces on the crystal morphology of urea crystal are investigated from first principle using the Hartman-Perdok approach. A procedure has been developed to find out surface termination when a slab of $d_{hkl}$ thickness is created having $\{hkl\}$ orientation. Slice energy of different habit faces have been computed and the attachment energy was calculated. The crystal growth morphology of the urea crystal has been drawn on the basis of computed attachment energies of different habit faces. The calculated crystal growth morphology depends on the functional employed for computing the attachment energies. The relaxed morphology obtained on the basis of attachment energy using Hartree-Fock method and B3PW density functional theory give a fairly good agreement with the as grown morphology from vapour phase while morphology computed using B3LYP functional underestimated $\{111\}$ facet and predicted the appearances of $\{101\}$ face. Polar solvent effects on the growth morphology were discussed particularly in the context of different atomic environments of $\{111\}$ and $\{\bar{1}\bar{1}\bar{1}\}$ faces.

[*]**Email: mksingh@cat.ernet.in**

# 1. INTRUDUCTION

The study and engineering of crystal facet have attracted immense interest of artists and crystal grower community since at least the Bronze Age [1]. The studies of crystal growth morphology are important for its device application, in which only some particular facet are usable and for pharmaceutical industries. Although significant efforts have been made for the last few decades, to predict precisely the growth morphology of crystals, however, it still remains a challenging task till this date. Crystals reveal a large variety of shapes, depending on the chemical composition and the structure of crystals, and the growth conditions. The shape of the crystals has a direct impact on the separation efficiency and the stability of crystalline chemicals, the bioavailability and the effective delivery of drugs, etc. Knowledge of the growth habits and morphological properties of the molecular crystals is of crucial importance in understanding and exploiting many of their physico-chemical properties. Calculating the crystal growth morphology has diverse application ranging from drug design [2] to explosives [3] and the inverse gas chromatography data [4].

The difficulty in understanding the crystal growth phenomena arise due to complex relationships between different processes, which take place on very large time and length scales, particularly in the case of industrial processes. The processes involved during the crystal growth are highly complex, involving many-body interactions for accurate description of the problem. The usefulness of the attempts to compute and predict the growth morphologies of crystals arises for the following reason. Crystals grown under a specific set of conditions (temperature, pressure, chemical composition, impurities, pH values, etc.) tend to have a specific morphology at least in the sense of statistics. In other words, if one measure the relative areas of the various habit faces for a substantial number of fairly large crystals of a given species, the ratios of these numbers are likely to be quite unique, with only a small scatter around the averages ratios. This set of numbers is indeed a very valuable information Nature have given to us about the mechanism it adopts for the crystal growth. The clue is not only valuable qualitatively; it is also very substantial quantitatively because a crystal usually has a fairly large number of habit faces. A theory which can predict this entire set of numbers correctly (within the statistical error) provide a good insight into what really happens in the interfacial layer between the growing crystal and the mother nutrient. Even more importantly, success in this venture gives insight into how one should choose the growth parameters to achieve a desired result.

Attempts have been made to relate crystal morphology to internal structure at the molecular level. Bravais-Friedel-Donnay-Harker (BFDH) [5] has made the first attempts among them. Taking solely the geometry of crystal lattices into account, Bravais et al. assumed that the relative growth rates $R_{hkl}$ of the crystal faces is inversely proportional to the inter-planar distance $d_{hkl}$. No account was taken of the chemical nature of the crystal or the molecular packing arrangement. The justification of the BFDH model is based on the fact that large inter-molecular interaction that leads to large slice energy has been resulted due to thicker inter-planar lattice spacing. The slice and attachment energy of a given face is complementary to each other and the growth rate is directly proportional to attachment energy. These arguments established the BFDH model. In spite of the success of BFDH theories in some cases, discrepancies between the observed and theoretically predicted morphology occurred, particularly in the case where crystals are either grown from solution or melt. By means of their periodic bond chain (PBC) theory, Hartman and Perdok [6] attempted to quantify crystal morphology in terms of the interaction energy between crystallising units. They used the assumption proposed by Born [7] that surface energy is directly related to the chemical bond energies. Hartmann and Benema [8] introduced the proportionality relation between the growth rate of the flat-face (F-face) and its attachment energy and showed its validity is limited to low super-saturations when the birth-and-spread and Burton-Cabrera-Frank (BCF) [9] model dominate the crystal growth mechanism. The attachment energy ($E_{att}$) is defined as the energy per mol of molecule released when a new layer having thickness $d_{hkl}$ is attached to the surface of the crystal. Attachment energy is a measure of the growth rate normal to a face, so that faces which have higher attachment energy grow faster and have lower morphological importance (M.I.). Using the Hartman-Perdok theory (HP) it is possible to compute the facets (*hkl*), which will dominate the morphology from known crystal structure. Although HP theory does not take into account of possible reconstruction or relaxation at the surface of habit face but is still quite successfully predict the crystal morphology. While cutting a slab of a given orientation (*hkl*), many surface terminations are possible, but the

HP theory does not take care of the effect of surface termination on the attachment energy. In most of the cases, F-face have slowest growth rate compared with step (S) and kink (K) faces and hence largest M.I. This is the reason why crystal faces mostly bounded by F-faces rather than S or K-faces.

The crystal growth morphology computed on the basis of its internal structure is best compared with the crystals which are obtained from sublimation method. The crystal morphology obtained from solution growth method, the effect of interactions between solute and solvents molecules at the various crystal interfaces may have a dominant role on the shape of the crystal. Habit modifications due to solvent were first studied by Wells [10], who explained the habit modification on the basis of preferential adsorption of solvent molecules on a particular faces, which delay the growth rate of these faces.

In the light of above discussed model for crystal growth phenomena, a fully *ab initio* calculation has been made for calculating the crystal growth morphology. In this reported work, urea crystal has been studied since it is an important commodity chemical used in the fertilizer industries and it also serves a better crystal compared to KDP for its non linear optical (NLO) coefficient [11]. The crystal growth morphology of urea crystal is well studied [12-15] due to its comparatively simple molecular structure. Docherty et al. [12] have calculated the atomic charge using *ab initio* Hartree-Fock method for monomers and dimmers of the urea molecule and also performed a full 3D calculation using periodic boundary calculation to derive atomic charge in unit cell. Boek et al. [13] were calculated the charge densities from *ab initio* Hartree-Fock method and decompose it into multipole moments to calculate the columbic contribution to the intermolecular interaction energy. However Docherty et al. [12] and Boek et al. [13] have not considered the effect of crystal faces like {101} and {200} and the relaxation of structure on surfaces while calculating the growth morphology. These faces correspond to large inter-planer spacing and required due consideration. On the other hand, Ashley et al. [14] have considered above facet while computing the growth morphology but they predicted the appearance of {101} face and overestimated the morphological importance of {200} and underestimated {111} face. All the above calculations were based on semi empirical force-field method where the atomic charges are derived either from empirical atomic charges or *ab initio* atomic charges. Symmetry of the slabs was not considered in these calculations. Moreover, the shift of origin is required to minimize the number of symmetry operators with finite translation components. Also in all earlier calculation, reconstruction of the molecules in slabs having orientation (*hkl*) was not taken into consideration while creating a slab and computing the relevant energy.

In this work, the growth morphology of crystalline urea derived from vacuum has been investigated from first principle method. The present paper describes the symmetry of the slab and possible reconstruction of the molecules in the slab within the frame work of Hatree-Fock and Density Functional theory. The physical slab must possess minimum energy configuration and zero dipole moment perpendicular to the slab. We have employed CRYSTAL03 [16] code to compute slice and attachment energies of habit faces having (*hkl*) orientation and $d_{hkl}$ thickness of urea crystal. We have taken care of reconstruction or relaxation at the surfaces of different habit face. Although we are aware that previously [12] CRYSTAL95 code has been used to obtained the atomic charges at each atomic site in a unit cell. The computed atomic charges were used to calculate lattice and attachment energies using semi empirical force field method to derive growth morphology of urea crystal.

The cohesive and attachment energies have been calculated using first principle method and the obtained energies was used in the empirical relation given by Hartmann and Benema [8] that growth rate of a given facet is proportional to its attachment energy while deriving the crystal growth morphology. Due to following inherent limitation of semi empirical force-field method, the first principle method was used for calculating the relevant energies. Some among them are following

(i) There is always an ambiguity exist while calculating the partial charges at atomic site since partial charges are not unique and it's strongly dependent on the methods [17, 18] used for derived the partial charges (ii) also the partial charges differs for bulk crystal and slices having orientation (*hkl*) (iii) The suitability of semi empirical method specially sort range potential is well studied for bulk while it is under question when it applied for surface and interface [19] (iv) Though the growth morphology is weakly depends on potential model [20], used for calculating the attachment energies, but the surfaces/slab created using first principle method is more accurate compared to its semi-empirical counterpart. The adsorption of impurity/solvent atoms on the surfaces created using the above is better studied by first principle method.

The present paper describes the first attempt to use *ab initio* method to compute the slice and attachment energy of habit faces. The Hamiltonian used for above calculations of energies is Hartree-Fock and Density functional theory. The relaxed slabs geometries have been obtained by performing Hartree-Fock and Hybrid Functional with density functional theory.

## 2. METHODOLOGY

The computer code used for computing slice and attachment energy of a habit faces having orientation $\{hkl\}$ is CRYSTAL03. An all electron Gaussian basis set are employed. A basis set employed for carbon, nitrogen and oxygen is $s(6)sp(2)sp(1)$ and for the hydrogen atom it is s(2)s(1) ( "6-21G" ). Initial geometry of the urea molecule is generated from Molden graphics code [21] and it is optimized using CRYSTAL03. Using space/point group symmetry of the urea crystal, we have computed the initial geometry of the asymmetric unit cell. The initial geometry is further subjected to geometry optimization using CRYSTAL03 code. Structural optimisations were performed by computing analytical energy gradients with respect to atomic fractional coordinates in the unit cell, and then standard conjugated gradients routine were used to carry out lowest energy configuration. To carry out full optimization i.e. cell parameters and atomic coordinates has been carried out in a two-step iterative process must be adopted. The Hartree-Fock and density functional method was employed to carry out bulk and slab geometry optimization. The cell parameters are optimised at fixed atomic positions. This was accomplished by a c-shell script called 'BILLY' [22]. Appropriate modification of the script has been done to run under parallel computing environment on Linux cluster. Atomic positions were being relaxed by keeping the lattice parameters at the previously optimum values. The above process is repeated till convergence is attained, i.e. cell constants satisfied the convergence criteria on nuclear forces are satisfied.

In order to correct the basis set superposition error (BSSE) in molecular crystals like urea; a counterpoise (CP) method of Boys and Bernardi, [23] taking into account the distortion of the reagents described elsewhere [24] was used and which is available with CRYSTAL03 code. In order to compute cohesive energy of bulk and slice having orientation parallel to (*hkl*), several Hamiltonian like Hartree-Fock and Density functional theory were studied. The B3LYP functional was chosen because of its success to molecule crystals [25-28]. The Becke's three-parameter functional [29]

$$E_{xc} = 0.50*(E_x^{LDA} + 0.90*E_x^{BECKE}) + 0.50*E_x^{HF} + 0.19*E_c^{VWN} + 0.81*E_c^{LYP} \quad …(1)$$

B3LYP/B3PW in CRYSTAL03 code is based on the 'exact' form due to Vosko-Wilk-Nusair [30] correlation potential.

The intermolecular binding (cohesive) energy per mol of molecule is given by:

$$E_{cohesive} = \frac{E_{crystal}}{n} - E_{molecule} \quad …(2)$$

Where $E_{crystal}$ is the total energy per unit cell and n is the number of molecule in a unit cell, $E_{molecule}$ is the intra molecular energy. We should note that for a stable crystal configuration, $E_{crystal} < 0$. In order to compute cohesive energy of a crystal, we have first compute $E_{crystal}$ and then molecular calculations were performed to compute intra molecular energy in crystal/slab. Notice the difference with respect to the usual binding energy, in which the reference corresponds to isolated atoms. We define cohesive energy according to equation (2) because of the fact that near the room temperature, integrity of urea molecules was maintained. The cohesive energy of crystal can be written in term of slice and attachment energy as follow:

$$E_{cohesive} = E_{slice}(hkl) + E_{att}(hkl) + E_{att}(\overline{hkl}) \quad …(3)$$

where $E_{slice}(hkl)$ is slice energy per mol of molecule. Each slice has two surfaces and $E_{att}$ is the corresponding attachment energy per mol of molecule. To compute pure crystals morphology drawn from vapour phase, interactions with the environment (solvent, impurities) should be neglected.

The polyhedron which is bounded by most stable faces i.e. slowest growth rate is called the 'theoretical form' of the crystal. The theoretical form of the crystal which is derived from the relative growth rates of the various low-index faces. The growth rate is given by:

$$R \propto E_{att} \qquad \qquad \ldots(4)$$

and

$$R_{rel} = \frac{R_i}{R_j} = \frac{E_{att}(i)}{E_{att}(j)} \qquad \qquad \ldots(5)$$

Where any two arbitrary faces of the crystal is denoted by i and j, which is usually the F-faces. Equation (4) can be applicable to some extend to predict growth morphology of crystals from the vapour and main habit controlling factors can depend on many factors like surface energy, inter-planer displacing etc [31]. The morphological importance for habit face is inversely proportional to growth rate i.e., fast growing facet do not appear in the crystal morphology which at the expense of slow growing facet. Prywer et al. [32] has derived an expression in which they shown that the size of a given face not only depends on growth rate of the given facet but also on the growth rate of neighbouring faces and their interfacial angles. The have shown that under some conditions faster growing face compared to its neighbouring faces has higher morphological importance (M.I.). This leads to a conclusion that the M.I. of a given face need not necessarily proportional to its growth rate.

  To compute the slice energy and attachment energy, a slab having two ideal semi-crystals has been created, each limited by an ideal surface which was parallel to the given plane (*hkl*) and thickness $d_{hkl}$ of the 3D lattice. The slab is characterised by its orientation, the label of the surface layer and number of atomic layers in the slab. The program itself checked the charge neutrality of the slab. The origin is moved in order to minimize the number of symmetry operators with finite translation components. This is required before cutting a slab from 3-D structure. The slab should possess zero dipole moment perpendicular to slab and it does possess minimum energy state. While cutting a slab for a given orientation (*hkl*), often more than one surface termination is possible. To determine the surface termination of a given slab of orientation, we follow above mentioned criterion. This will fix the surface layer of a given slab having orientation. The number of layer in a slab is determined by its inter-planer spacing distance. Each semi-crystal preserves 2-D periodicity parallel to the selected face, but loses all symmetry elements, which involve displacements in a perpendicular direction. Hence, the ideal surface may undergo relaxation, without loss of transactional symmetry, or exhibit partial reconstruction. The MOLDEN Graphics [21] package was used for calculating the required displacements of atoms in 2D unit cell. XCRYSDEN [33, 34] was employed to reconstruct the fragmented molecules in a slab having (*hkl*) orientation.

## 3. RESULTS AND DISCUSSION

The cohesive energy of urea crystal ($E_{cohesive}$) was computed using all electron basis set. Both Hartree-Fock and density functional theory was employed for computing cohesive energy for bulk and slab crystal. The hybrid functional was used for all density functional calculations. The calculated cohesive energy of urea crystal is presented in Table 1 at zero Kelvin temperature. The reported value of experimental sublimation enthalpy at room temperature is -21.0 kcal mol$^{-1}$ [35]. The cohesive energy reported in the literature are derived from the observed enthalpy of sublimation given by

$$E_L = \Delta H - 2RT \qquad \qquad \ldots(6)$$

where $\Delta H$ is the experimental enthalpy of sublimation and the last term of Equation (6) represents an approximate correction for the difference between the gas phase enthalpy for an ideal gas ( PV + 3RT) and the estimated vibrational contribution to the crystal enthalpy 6RT [36]. The cohesive energy given by above expression corresponds to an idealized potential energy at zero Kelvin temperature.

Using Equation (6) the extrapolated enthalpy of sublimation at zero Kelvin temperature is -22.15 kcal mol$^{-1}$. The cohesive energy obtained using Hartree-Fock method was a close agreement with the extrapolated value of experimental sublimation enthalpy and also in agreement with the Docherty et al. [12] calculation while Boek et al. [13] have overestimated the cohesive energy. One should note that Hartree-Fock method underestimated the cohesive energy due to lack of electrons correlation functional. On the other hand DFT calculation using B3LYP and B3PW functional overestimated the cohesive energy of urea crystal, which is due to the fact that correlation energy is less pronounced in molecular crystal. We have taken care of BSSE estimate while computing cohesive energy for urea crystal.

Fig. 1 shows the cleavage plane parallel to {200}. There exist eight different surface terminations for this plane but only the shown physical surface termination possesses lowest energy configuration state. The corresponding surface termination for {200} plane is hydrogen (H) and it has seven atomic layer within $d_{200}$ thickness. The origin was shifted to 0.056 Å for maximizing the symmetry of {200} plane. Fig. 2-5 shows the cleavage planes parallel to {110}, {101}, {001} and {111}. Table 4 shows the details results for construction of the various cleavage planes.

The slice energy for the {200}, {110}, {101}, {001} and {111} faces was computed, and, using Equation (3), the attachment energy $E_{att}$ was deduced are listed in Table 2 for the un-relaxed and relaxed structure. The above energies were calculated using Hartree-Fock and density functional method. All considered faces having low index plane and have largest inter-planer spacing. The {200} faces was considered rather than {100} because {200} face represented the irreducible growth slice for {100} face [10]. It is clear from Table 2 that, in all cases, the attachment energy increases upon relaxation. The relaxed attachment energy calculated using Hartree-Fock method increases by 9.5% to 18.75% while the relaxed attachment energy computed using density functional method increases by 7% to 14% depending on the facet under consideration. This clearly shows that structural relaxation should be considered in computational investigation of surface properties. Our studies shows that structural relaxation in molecular crystal is more pronounce compared to earlier calculation by Ashley et al. [14] using semi empirical force-field method. Ashley et al. [14] have calculated the un-relaxed and relaxed attachment energy for various habit faces and it is shown in Table 3. The un-relaxed attachment energies calculated by Docherty et al. [12] and Boek et al. [13] are presented in Table 3 for comparisons purpose. Following are the relative morphological importance (M.I.) of the different habit faces for un-relaxed and relaxed structure of different Hamiltonian.

1. Hartree-Fock

(a) Un-relaxed structure:

    M.I. {200} > M.I. {110} > M.I. {001} > M.I. {111} > M.I. {101}

(a) Relaxed structure:

    M.I. {200} > M.I. {110} > M.I. {001} > M.I. {111} > M.I. {101}

2. B3LYP/B3PW

(a) Un-relaxed structure:

    M.I. {110} > M.I. {200} > M.I. {001} > M.I. {111} > M.I. {101}

(a) Relaxed structure:

    M.I. {110} > M.I. {200} > M.I. {001} > M.I. {111} > M.I. {101}

Both B3LYP and B3PW functional predict the same morphological pattern for various habit faces, though the calculated slice and attachment energies differ considerably. Hartree-Fock method predicts highest morphological importance for {200} face but B3LYP and B3PW both predicted {110} has highest M.I. The morphology predicted using HF does not show appearance of {101} form. Fig. 6 shows the computed (a) un-relaxed (b) relaxed growth morphology of urea using Hartree-Fock method. The relaxed morphology computed using HF method is in agreement with scanning electron micrographs, shown in ref. [12], both our computed morphology and the experimental morphology have {110}, {001}, {111} or {$\bar{1}\bar{1}\bar{1}$} faces in crystalline urea. On the other hand our predicted morphology shows the appearance of {200} faces. It should be notes that {200} and {111} faces become more prominent after allowing the structural relaxation but {110} show less morphological importance after structural relaxation.

    Docherty et al. [12] predicted highest M.I. for {110} face while lowest for {111} face. They do not consider {200} and {101} faces and no consideration have been made for structural relaxation while calculating the growth morphology. Our computed growth

morphology using B3LYP and B3PW predict the same tends. However Ashley et al. [14] have calculated highest M.I. for {110} face and lowest for {111}.Their predicted growth morphology does not show the {111} face, which is present in the experimental growth morphology.

The morphology predicted using B3LYP and B3PW methods shows the highest M.I. for {110} face which is consistent with experimental morphology [12]. Figure 7 shows the computed (a) un-relaxed (b) relaxed growth morphology of urea using B3LYP method and Figure 8 shows the computed (a) un-relaxed (b) relaxed growth morphology of urea using B3PW method. The un-relaxed morphology computed using B3LYP and B3PW shows same forms as HF but the relaxed morphology clearly shows the appearance of {101} face. The relaxed morphology computed using B3PW functional predicted lower M.I. for {101} face compared to B3LYP. The un-relaxed/relaxed morphology predicted using B3PW method has also good agreement with experimental result [12].

Brunsteiner et al. [20] have been reported that the crystal growth morphology is weakly depend on the potential model chosen for computing the bulk cohesive and attachment energy however they found that the value of the lattice and attachment energy considerably depend on the potential model. We have employed three potential model for calculating the growth morphology of urea crystal using an *ab initio* method, and found that not only the bulk cohesive and attachment energies depend on the chosen potential but the morphology of the urea crystal also depend on the potential model. Also the surfaces/slab created using first principle method is more accurate compared to its semi-empirical counterpart. We have taken care of symmetry of the slabs. The adsorption of impurity/solvent atoms on the surfaces created using the above is better studied by first principle method.

An attempt has been made to compute the attachment energy of {111} and {$\bar{1}\bar{1}\bar{1}$} face before and after structural relaxation but we could not find difference between the attachment energy of {111} or {$\bar{1}\bar{1}\bar{1}$} even after relaxation have been performed.

The structural polarity can be resulted due to the unequal growth of {$hkl$} and {$\overline{hkl}$} facet, which leads to a polar morphology. Figure 1-5 revealed that slab having orientation {200}, {110}, {101} and {001} have symmetric atomic environment along z-axis but {111} and {$\bar{1}\bar{1}\bar{1}$} have different atomic environment. From the structural point of view it should be noted that {200} and {$\bar{2}00$} have same growth rate. On the same reasoning {110} and {$\bar{1}\bar{1}0$}, {001} and {$00\bar{1}$}, {101} and {$\bar{1}0\bar{1}$} have the same growth rate. The growth rates of {111} and {$\bar{1}\bar{1}\bar{1}$} can differ, not for structural, but rather for environmental reasons. For example, the polarity of the face may determine phenomena such as salvation, as also the identity and orientation of the growth of atoms which may tend to attach to such a face during growth of the crystal.

## 4. CONCLUSION

In the present paper a first principle calculation has been performed to compute the slice and attachment energy for various habit faces of crystalline urea. A method was developed to create a slab and the symmetry of the created slab was taken into consideration. Reconstruction/relaxation has been performed to all habit faces before attempted to compute the slice energy. In following section, one has seen the good agreement between our relaxed computed morphology using Hartree-Fock and B3PW method and the observed morphology obtained from sublimation. The results of our study and others [37, 38] published calculations on crystal morphology indicated that for crystals grown from low super-saturations, the effects external factor like fluid motion, super saturation etc during crystallisation process on the morphology of the crystal are negligible. We have applied the above *ab initio* method to predict the morphology of beta-succinic acid and other molecular crystals like antracine and find excellent agreement with as grown morphology.

      In the reported work, we have employed three Hamiltonian for calculating the growth morphology of urea crystal from first principle method using periodic boundary condition, and found that not only the cohesive energy for bulk crystal and attachment energies for different habit faces of urea crystal strongly depend on the chosen Hamiltonian but the computed morphology of the urea crystal also depend on the Hamiltonian. The surface created using first principle method is more accurate compared to semi-empirical method because of atomic charge depends on the surface structure. We have taken care of symmetry of the slab while creating the slab. The adsorption of

impurity/solvent atoms on the surfaces created using the above is better studied by first principle method. The effect of impurity/solvent on the each habit faces of the crystal can be analyzed, which are cleaved from a pure crystal.

In the work reported here the slice energy is defined as the energy released per mol of growth unit when a two dimensional infinite slab of orientation {*hkl*} having $d_{hkl}$ thickness are formed. These slab will further adsorbed on the {*hkl*} faces of the crystal for its growth. According to this model, we will always get a cleavage plane parallel to {*hkl*} face. It is interesting to note from ref. [39] that monomers or dimmers of molecules adsorbed during the crystal growth. The crystal growth model proposed by Burton et al. [9] suggested that screw dislocation on the crystal face provide step (S) and a layered mechanism is responsible for the origin of flat (F) facet. Motivated from above studies we have developed a new model for computing the attachment energy of growth unit as monomers or dimmers of molecules. The free standing monomers or dimmers of molecules in solution/vacuum diffuses towards a given {*hkl*} face and when its reaches very close to surface, the growth unit tries to readjust its atomic position due to presence of surface environment. The growth unit also tries to reorient itself according to the atomic/molecular environment of the surface before it adsorbed to {*hkl*} face of the crystal. For calculating the crystal growth morphology of urea and other molecular crystals, the proposed model has been applied. Details will be published in due course.

## ACKNOWLEDGMENTS

The author is grateful for the support and motivation received from Dr. V.K. Wadhawan. The author feels much indebted to Dr. Arup Banerjee and Dr. V.S. Tiwari for valuable discussions and a critical reading of this manuscript.

## Captions of tables:

Table 1. The experimental and computed cohesive energy of urea crystal using various Hamiltonian.

Table 2. The calculated un-relaxed and relaxed slice and attachment energy of different habit faces of urea crystal using *ab initio* method for various Hamiltonian.

Table 3. The computed un-relaxed and relaxed attachment energy of different habit faces of urea crystal by others, using semi empirical force-field method.

Table 4. The construction of slab for various cleavage planes of urea crystal.

**Captions of figures:**

Fig. 1. The relaxed structure of {200} plane of urea crystal.

Fig. 2. The relaxed structure of {110} plane of urea crystal.

Fig. 3. The relaxed structure of {101} plane of urea crystal.

Fig. 4. The relaxed structure of {001} plane of urea crystal.

Fig. 5. The relaxed structure of {111} plane of urea crystal.

Fig.6. Calculated crystal growth morphology using Hartree-Fock method, (a) un-relaxed and (b) relaxed of crystalline urea.

Fig.7. Calculated crystal growth morphology using B3LYP Hamiltonian, (a) un-relaxed and (b) relaxed of crystalline urea.

Fig.8. Calculated crystal growth morphology using B3PWHamiltonian, (a) un-relaxed and (b) relaxed of crystalline urea.

| Hamiltonian | Cohesive Energy per molecule ($E_{cohesive}$) (kcal/mol) | Experimental sublimation enthalpy (extrapolated at zero Kelvin) (kcal/mol) [35] | Others calculation using semi empirical force-field method (kcal/mol) [12] | [13] |
|---|---|---|---|---|
| Hartree-Fock | -21.290 | | | |
| B3LYP | -27.004 | -22.15 | -22.6 | -26.2 |
| B3PW | -27.004 | | | |

Table 1

| Hamiltonian/ Functional | Forms {$hkl$} | Un-relaxed Energy (kcal/mol) | | relaxed Energy (kcal/mol) | |
|---|---|---|---|---|---|
| | | $E_{slice}$ | $E_{att}$ | $E_{slice}$ | $E_{att}$ |
| Hartree-Fock | {200} | -13.33 | -3.98 | -12.12 | -4.58 |
| | {110} | -13.28 | -4.00 | -11.78 | -4.75 |
| | {101} | -6.69 | -7.30 | -5.31 | -7.99 |
| | {001} | -11.21 | -5.08 | -9.99 | -5.65 |
| | {111} | -9.05 | -6.12 | -7.40 | -6.95 |
| B3LYP | {200} | -14.88 | -6.06 | -13.88 | -6.56 |
| | {110} | -16.84 | -5.08 | -15.42 | -5.79 |
| | {101} | -8.70 | -9.15 | -7.40 | -9.80 |
| | {001} | -13.37 | -6.82 | -12.12 | -7.44 |
| | {111} | -10.73 | -8.14 | -9.11 | -8.95 |
| B3PW | {200} | -14.05 | -6.48 | -13.02 | -6.99 |
| | {110} | -15.09 | -5.96 | -13.63 | -6.69 |
| | {101} | -7.26 | -9.87 | -5.92 | -10.54 |
| | {001} | -12.31 | -7.35 | -11.05 | -7.97 |
| | {111} | -9.31 | -8.85 | -7.71 | -9.64 |

Table 2

| Forms {hkl} | Un-relaxed Energy by others (kcal/mol) | | | Relaxed Energy by other [14] (kcal/mol) |
|---|---|---|---|---|
| | $E_{att}$[12] | $E_{att}$[13] | $E_{att}$[14] | $E_{att}$ |
| {200} | ---- | ----- | -6.63 | -6.56 |
| {110} | -4.50 | -4.30 | -4.86 | -4.98 |
| {101} | ---- | ---- | -8.26 | -8.12 |
| {001} | -5.20 | -5.82 | -5.92 | -5.94 |
| {111} | -6.00 | -9.31 | -9.00 | -9.28 |

Table 3

| Form {hkl} | No. of possible termination | Surface | | No. of Atomic layer in the slab | Origin shift to maximize symmetry (Å) |
|---|---|---|---|---|---|
| | | Atom | label | | |
| {200} | 2 | H | 10, 12 | 7 | 0.056 |
| {110} | 2 | C, N | 2, 7 | 7 | 0.388 |
| {101} | 1 | O | 4 | 16 | 0.250 |
| {001} | 2 | C, H | 2, 10 | 10 | -0.393 |
| {111} | 1 | N | 5 | 13 | 0.144 |

Table 4

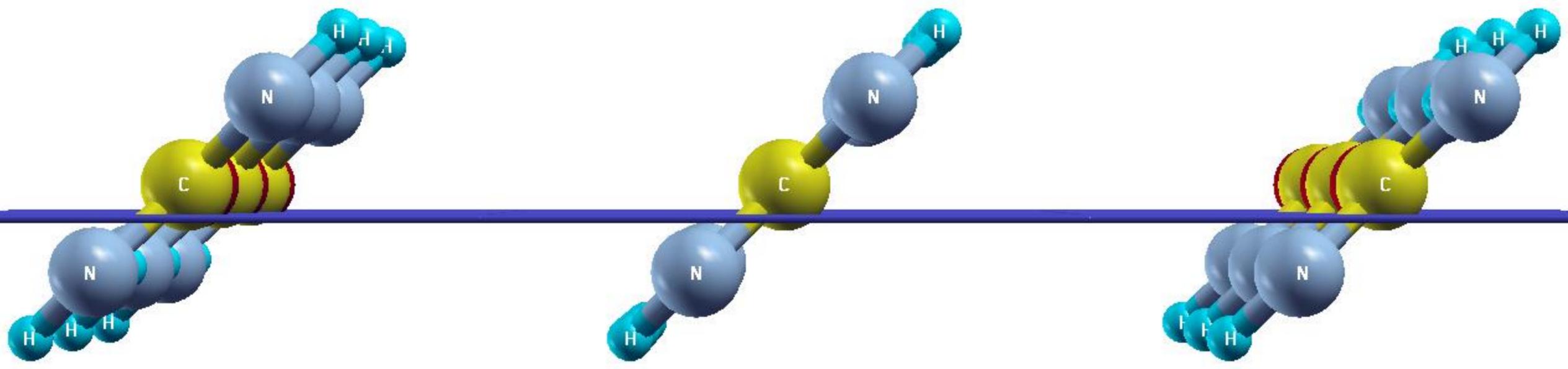
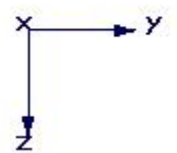

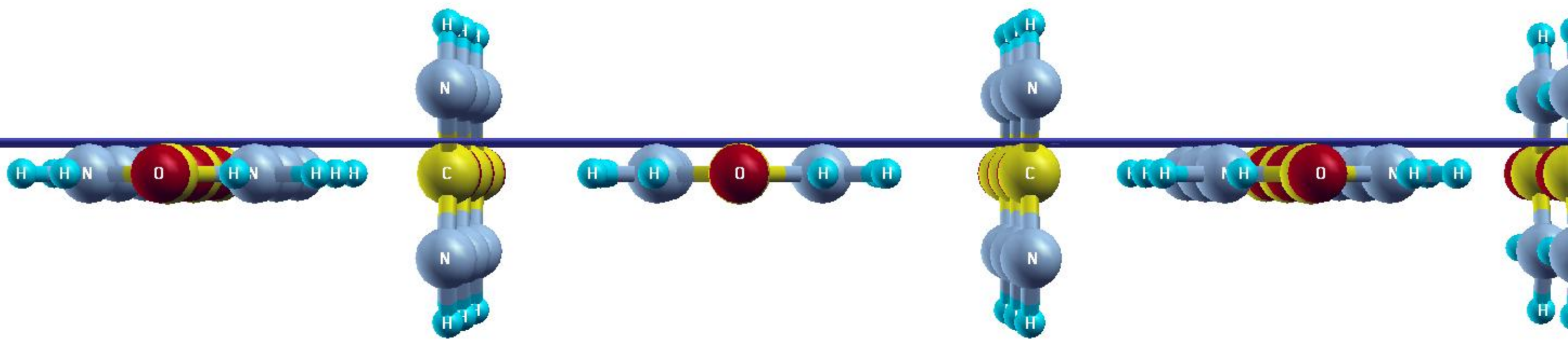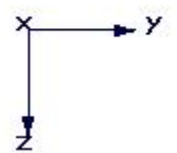

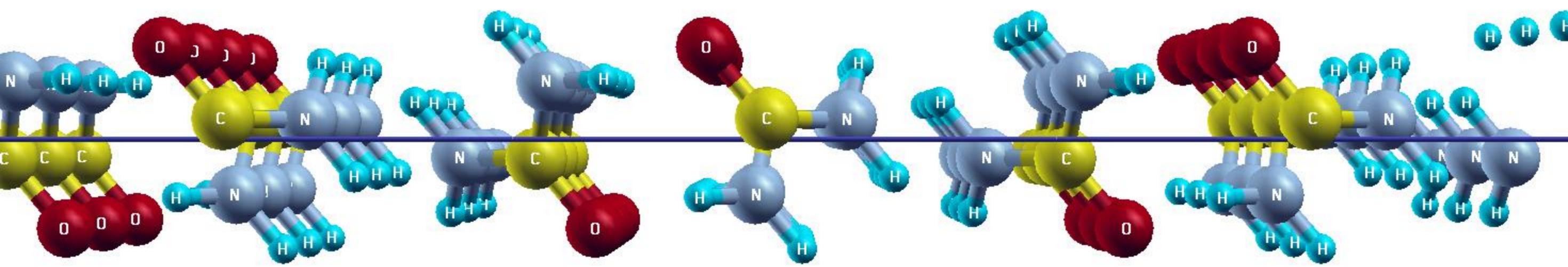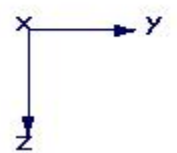

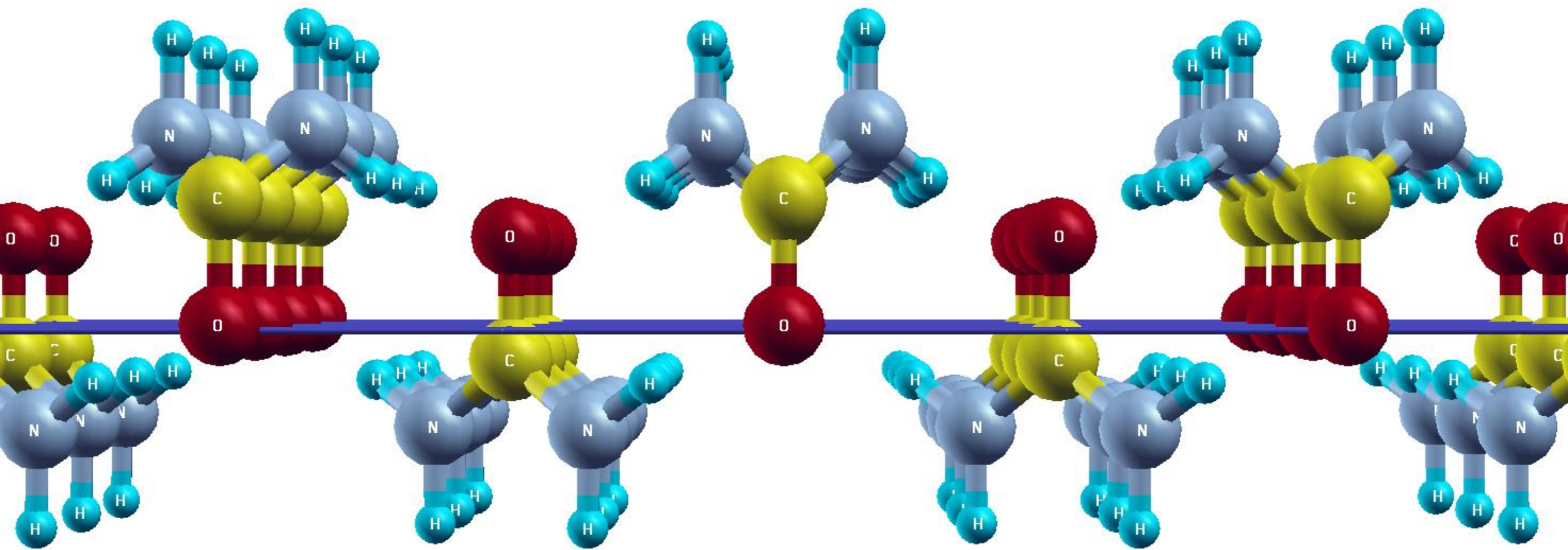
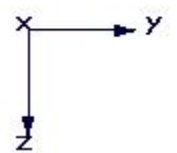

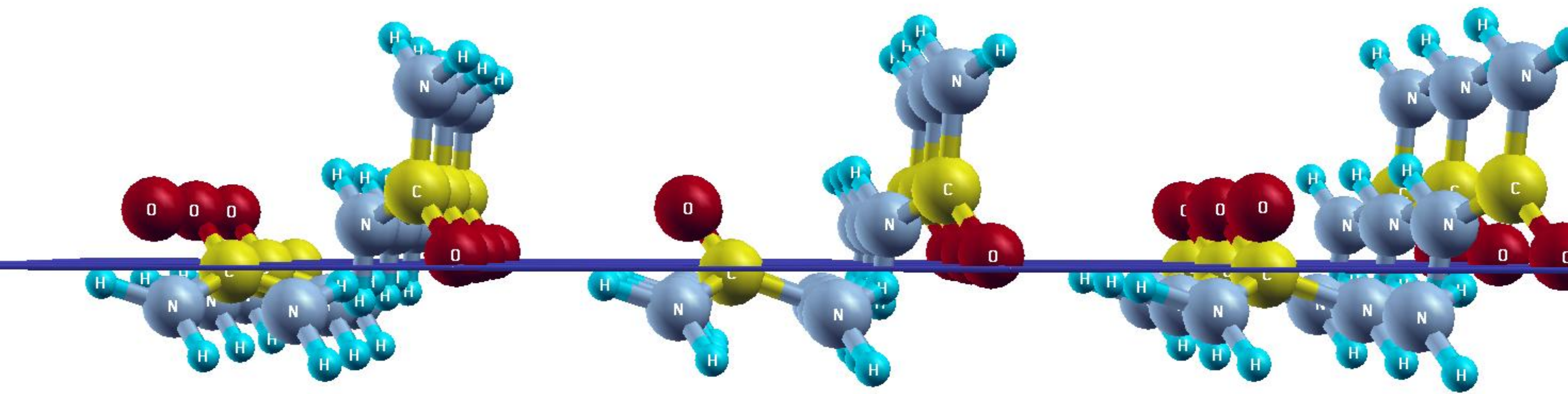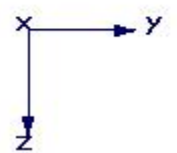

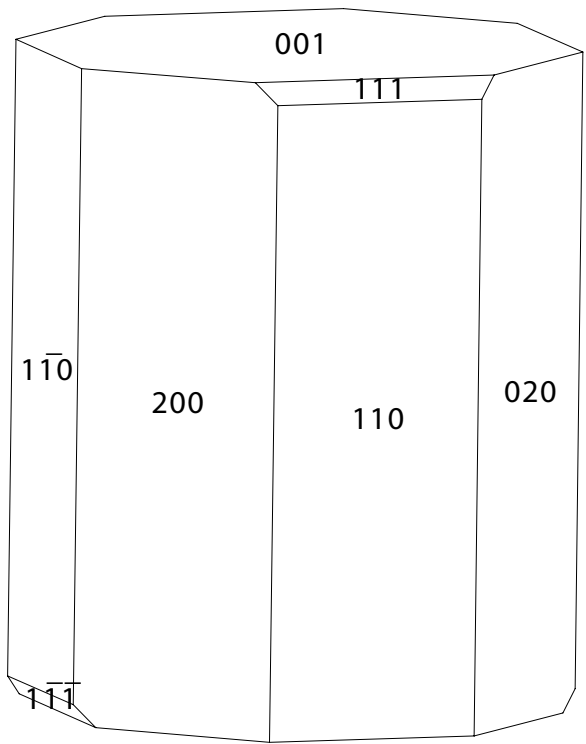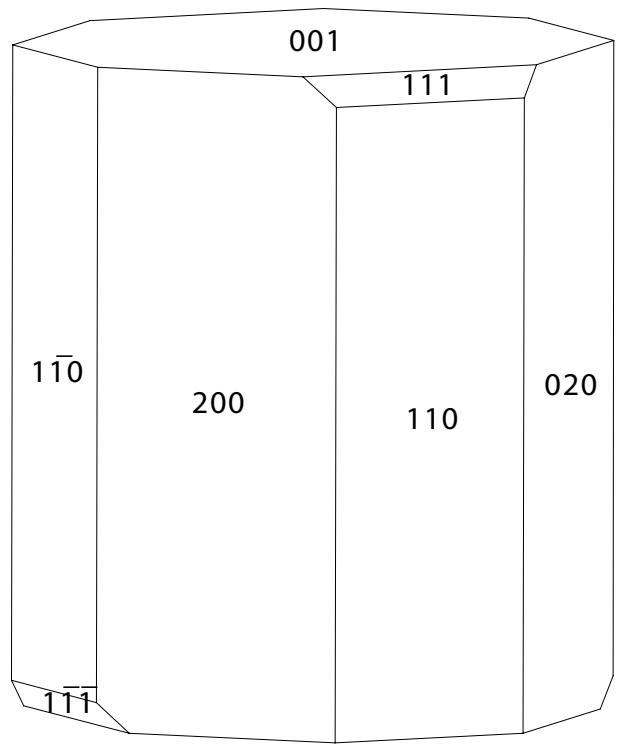

(a)                                      (b)

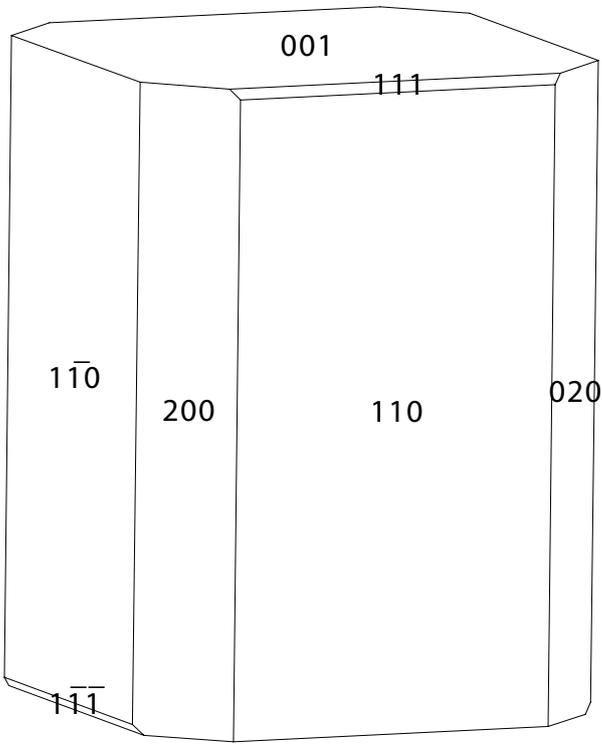 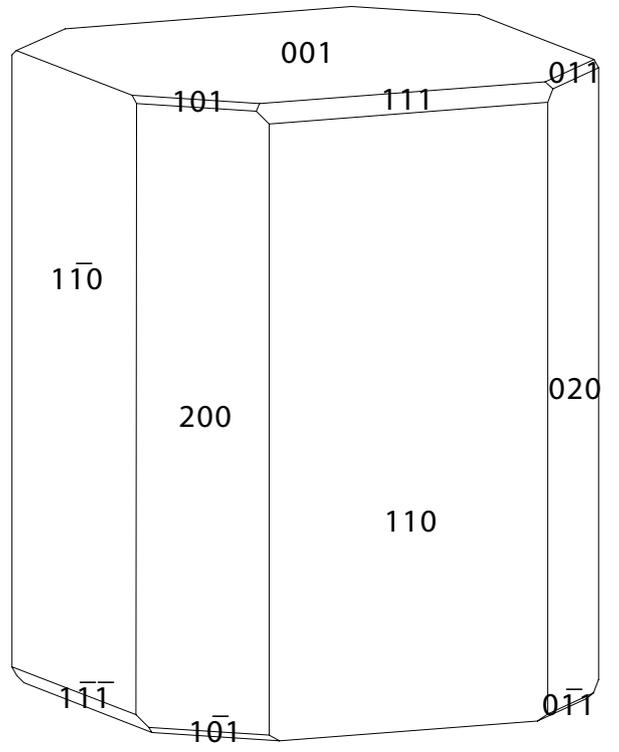

(a) (b)

(a)

(b)